 \documentclass[pra,twocolumn,lengthcheck,superscriptaddress,showpacs,preprintnumbers,amsmath,amssymb]{revtex4}

 \usepackage{graphicx}
 \usepackage{dcolumn}
 \usepackage{bm}

 \usepackage[T1]{fontenc}
 \usepackage{ae,aecompl}

\begin{document}

 \title{Attractive ultracold bosons in a necklace  optical lattice}

 \author{P. Buonsante}
 \affiliation{Dipartimento di Fisica, Politecnico di Torino Corso Duca degli Abruzzi 24, I-10129 Torino (ITALIA)}%
 \author{V. Penna}
 \affiliation{Dipartimento di Fisica, Politecnico di Torino Corso Duca degli Abruzzi 24, I-10129 Torino (ITALIA)}%
 \author{A. Vezzani}
 \affiliation{Dipartimento di Fisica, Universit\`a degli Studi di Parma and C.N.R.-I.N.F.M., Parco Area delle Scienze 7/a, I-43100 Parma (ITALIA)}
\date{\today}

 \begin{abstract}
We study the ground state properties of the Bose-Hubbard model with attractive interactions on a $M$-site one-dimensional  periodic --- necklace-like --- lattice, whose experimental realization in terms of ultracold atoms is promised by a recently proposed optical trapping scheme, as well as by the  control over the atomic interactions and tunneling amplitudes granted by well-established optical techniques. We compare the properties of the quantum model to a semiclassical picture based on a number-conserving su$(M)$ coherent state, which results into a set of modified discrete nonlinear Schr\"odinger equations. We show that, owing to the presence of a correction factor ensuing from number conservation, the ground-state solution to these equations provides a remarkably satisfactory description of its quantum counterpart not only --- as expected --- in the weak-interaction, {\it superfluid regime}, but even in the deeply quantum regime of large interactions and possibly small populations. In particular, we show that in this regime, the delocalized, Schr\"odinger-cat-like quantum ground state can be seen as a coherent quantum superposition of the localized, symmetry-breaking ground-state of the variational approach. We also show that, depending on the hopping to interaction ratio,  three regimes can be recognized both in the semiclassical and quantum picture of the system.
 \end{abstract}


 \maketitle
\section{Overview}
\label{S:intro}
Owing to the impressive progress in experimental techniques, ultracold neutral atoms trapped in optical lattices are nowadays widely recognized as a versatile toolbox bringing into reality ideal models of condensed matter physics \cite{A:Toolbox}.
A prominent example in this respect is no doubt the Bose-Hubbard model, originally introduced to sketch the physics of superfluid helium in porous media \cite{A:Fisher} and subsequently shown to be realizable in terms of optically trapped  ultracold Bosonic atoms \cite{A:Jaksch,A:Greiner02}. This  model,  describing interacting Bosonic particles hopping across the $M$ sites of a discrete structure, is characterized  by a Hamiltonian of the form
\begin{equation}
\label{E:BHH}
 H = \frac{U}{2} \sum_{m=1}^M n_m (n_m-1) - T \sum_{m,m'=1}^M J_{m m'} a^\dag_m a_{m'}
\end{equation}
where $a_m$, $a_m^\dag$ and $n_m = a^\dag_m a_m$ are on-site Bosonic operators, $U$ measures the strength of the (on-site) boson-boson interaction, $T$ is the hopping amplitude across neighboring sites and $J$ is the so-called adjacency matrix, describing the lattice topology. Its generic entry, $J_{m m'}$, equals 1 if the sites $m$ and $m'$ are adjacent, and 0 otherwise.
Parameters  $U$ and $T$ are directly related to well defined experimental quantities, i.e. the scattering length of the Bosonic atoms and the intensity of the laser beams giving rise to the optical lattice, respectively \cite{A:Jaksch}. The possibility of tuning the lattice strength over a wide range of values played a fundamental role in the experimental observation  \cite{A:Greiner02} of the superfluid-insulator quantum phase transition predicted for Hamiltonian (\ref{E:BHH}) in the case of repulsive interaction \cite{A:Fisher}. 
Further aspects of versatility of optically trapped ultracold atoms lie in the possibility of tuning the atomic scattering length, and hence the boson-boson interaction, via Feshbach resonances \cite{A:Theis} and realizing a great variety of different lattices, by means of a suitable laser beam setup \cite{A:Blakie,A:Santos}, possibly involving holographic elements \cite{A:Amico05}.

In this paper we are concerned with the case of attractive bosons in a
1D  periodic lattice, whose experimental realizability is made concrete
by a recently proposed optical trapping scheme \cite{A:Amico05} and by
the control over the sign and intensity of the boson-boson interaction
granted by Feshbach resonances \cite{A:Theis}. This symmetric, {\it necklace-like} geometry is
particularly interesting since, in the strong interaction limit, the
system ground state is a quantum superposition of macroscopic
spatially localized states, i.e. a so-called {\it Schr\"odinger-cat}
state \cite{N:cat}. Although the state-of-art experimental setups ensure a
remarkable control and insulation from the environment, the
realization of such deeply quantum states presents several
difficulties, which will be briefly discussed in the final part of
this work. Our main aim here is a study of the ground state properties of quantum Hamiltonian (\ref{E:BHH}), which we compare to the corresponding state of a suitable semiclassical approximation to the Bose-Hubbard model.

Before going into details, we mention that the attractive Bose-Hubbard model has been addressed by two recent works \cite{A:Dorignac2,A:Jack}. In Ref.~\cite{A:Dorignac2} a further nonlinear (cubic) repulsive term is introduced in Hamiltonian (\ref{E:BHH}) in order to prevent excessive localization for large values of the interaction. Such a term is ascribed to three body interactions coming into play at elevate atomic densities. However, experimental observations suggest that these interactions produce different phenomena resulting into a leak of condensate atoms rather than into a repulsive term  \cite{N:collapse}.
Hence, here we do not consider such a repulsive term, and assume that the total boson population is sufficiently low to rule out the above phenomena. In some respects our analysis is similar to that of Ref.~\cite{A:Jack}, whose results we compare to ours. Other recent results related to ours are found in Refs.~\cite{N:Kanamoto}, which also focus on annular geometries, yet in the absence of a lattice potential.
 Finally, we mention that the attractive Bose-Hubbard model~(\ref{E:BHH}) has received considerable attention as the quantum counterpart of  the discrete self-trapping (DST) or quantum discrete non-linear Schr\"odinger (DNLS) equations \cite{A:Scott86,A:Bernstein}, particularly in relation to the issues of discrete quantum solitons \cite{A:Scott94} and breathers \cite{A:D_Breathers,A:Fleurov}.

The layout of this paper is as follows. In Section \ref{S:Quantum} we rapidly review the key features of the Bose-Hubbard model (\ref{E:BHH}). Focusing on the special case of a periodic one-dimensional lattice comprising $M$ sites, we recall the main properties arising from translational invariance. 
In Section \ref{S:semicl} we describe a semiclassical variational approach to Hamiltonian  (\ref{E:BHH}) based on a number-conserving trial state. This approach results into a set of DNLS equations featuring a correction factor that makes them effective also in the deeply quantum regime characterized by large attractive interactions and small boson populations. Furthermore, we describe the three regimes that can be recognized for the ground state of the one-dimensional periodic lattice under investigation. A {\it Schr\"odinger-cat regime} for small values of the parameter $\tau= T/|U| ({\cal N}-1)$, where $\cal N$ is the total number of bosons in the system; a {\it soliton regime}, for intermediate values of $\tau$; a {\it superfluid regime} for large values of $\tau$. We show that the crossover between the Schr\"odinger-cat and soliton regimes occurs around a size-independent threshold $\tau_1$, while the critical point marking the boundary between the soliton and superfluid regimes, $\tau_2^{(M)}$, exhibits a clear dependence on the lattice size $M$, which becomes linear for large sizes. In Section \ref{S:results} we illustrate the effectiveness of the semiclassical description introduced in Section \ref{S:semicl}. In particular, we show that in the Schr\"odinge-cat regime, the delocalized ground state of the quantum model is very well described by a coherent superposition of the degenerate localized ground states of the semiclassical picture. Furthermore, we discuss the relevance of the semiclassical thresholds $\tau_1$ and $\tau_2^{(M)}$ in the quantum picture. In particular we show that the local observables considered in Ref. \cite{A:Jack} are not sensitive to size dependent threshold  $\tau_2^{(M)}$, which is related to the finiteness of the system, but rather to the size-independent threshold $\tau_1$, which is related to the discrete nature of the system. We also show that  $\tau_2^{(M)}$ is signaled by nonlocal observables.
In Appendix \ref{S:suM} we show that the variational trial state introduced in Section \ref{S:semicl} is actually a su$(M)$ coherent state, which is a further justification (?) for the effectiveness of the semiclassical picture. In Appendix \ref{S:pert} we show that in the Schr\"odinger-cat regime the localized ground state of the semiclassical picture coincides ---  at the first perturbative order in $\tau$ --- with a coherent superposition of the almost degenerate $M$ lowest states of Hamiltonian (\ref{E:BHH}).

\section{Homogeneous periodic 1D lattice}
\label{S:Quantum}
Let us now summarize the main properties of Hamiltonian~(\ref{E:BHH}).
First of all, we recall that the relevant Hilbert space is in principle infinite, and it is spanned by the Fock vectors generated by the action of the positive powers of the creation operators $a^\dag_m$ onto the vacuum state $|\Omega\rangle$, i.e. the state such that $a_m |\Omega\rangle =0$ for any $m =1,2,\ldots,M$. However, Hamiltonian~(\ref{E:BHH}) commutes with the total number operator,  $N = \sum_{m=1}^M n_m$, and hence its study can be restricted to the finite size eigenspaces of $N$. Henceforth, we assume to be in the eigenspace relevant to the total population $\cal N$, whose size is $d({\cal N},M)=\binom{{\cal N}+M-1}{\cal N}$.
As we mention, we are interested in the case of attractive bosons ($U<0$) on a 1D homogeneous periodic lattice.  Hence we particularize the adjacency matrix in Eq.~(\ref{E:BHH}) to
\begin{equation}
\label{E:homJ}
J_{m m'}=\delta_{m\, m'+1}+\delta_{m\, m'-1},
\end{equation} 
where, on a periodic lattice comprising $M$ sites, the site labels $0$ and $M+1$ correspond to sites $M$ and $1$, respectively. With this choice Hamiltonian~(\ref{E:BHH}) is clearly translationally invariant. Hence all of the relevant observables cannot break this symmetry, i.e. exhibit a dependence on the site label.
Owing to the translational invariance the eigenstates of $H$ can be labeled by a further quantum number. 
This is made quite evident by switching to the reciprocal space, i.e. by 
introducing the transformed operators 
\begin{equation}
\label{E:bk}
b_k = M^{-1/2} \sum_{m=1}^M e^{i 2 \pi k m/M} a_m,
\end{equation}
 making the kinetic term of Hamiltonian (\ref{E:BHH}) diagonal. Of course, the reciprocal space has the same size $d({\cal N},M)$ as the direct space, and it is spanned by the reciprocal Fock vectors, generated by the action of the powers of the $b^\dag_p$'s on $|\Omega\rangle$. Note indeed that $N=\sum_{p=1}^M b^\dag_p b_p$ and that $b_p |\Omega\rangle = 0$ for any $p$.
 In this representation the displacement operator changing $a_m$ into $a_{m+1}$ is simply 
\begin{equation}
\label{E:disp}
D = \exp(i \frac{2\pi}{M} P), \quad P = \sum_{p=1}^M p \, b^\dag_p b_p .
\end{equation}
Since $[D, H]=0$, the Hamiltonian can be studied separately in the $M$
eigenspaces of $D$. Note indeed, that $D^M={\mathbb I}$, which entails
that the eigenvalues of the displacement operator are the $M^{\rm th}$ roots of
unity. After noticing that  the reciprocal Fock vectors are 
clearly eigenstates of
$D$, the block diagonal structure of $H$ can be made evident by simply
arranging these states according to the relevant eigenvalue, which we denote
by $\cal D$.
A perhaps more convenient choice for the displacement
quantum number is the integer $k \in [0,\, M-1]$ such that  $ -i\log({\cal D})=
2\pi k/M $. In
particular, the ground state of the system is the lowest eigenstate of
the block relevant to ${\cal D}=1$ or, equivalently, $k = 0$.
Introducing a further integer label $j$ ranking the eigenstates of
of Hamiltonian (\ref{E:BHH}) according to their energy, we set
\begin{equation}
\label{E:Hspec}
D |\psi_j^{(k)}\rangle = e^{i\frac{2\pi}{M}k} |\psi_j^{(k)}\rangle ,\qquad H
|\psi_j^{(k)} \rangle = E_j^{(k)} |\psi_j^{(k)}\rangle 
\end{equation}
with $E_j^{(k)}\leq E_{j+1}^{(k)}$.
In particular, $|\psi_1^{(0)}\rangle $ and $E_1^{(0)}$ denote the ground state
of $H$ and its energy. We recall that this state is translationally 
invariant, while, more in general, all of the eigenstates of $H$ are
 delocalized. That is to say,  these eigenvalues cannot give rise to
 site-dependent expectation values of local operators, such as e.g. $n_m$.

This means that, at least in principle, the states $|\psi_1^{(k)}\rangle $ exhibit the features of the  celebrated {\it Schr\"odinger cat} state.  
Indeed, as it is well known, for $T \ll U$  the ground states of the $M$ blocks of $H$ are almost degenerate coherent superpositions of macroscopically localized states of the form 
\begin{equation}
\label{E:loc}
|\psi_1^{(k)}\rangle =  \sum_{m=1}^M \frac{e^{i\frac{2\pi}{M} m k }}{\sqrt M} |{\cal N}_m\rangle, \quad|{\cal N}_m\rangle = \frac{(a^\dag_m)^{\cal N}}{\sqrt{{\cal N}!}} |\Omega\rangle.
\end{equation}
 This can be quite easily seen by treating the kinetic term of
 Hamiltonian (\ref{E:BHH}) perturbatively  \cite{A:Bernstein}.
In Sec. \ref{S:results}  we show that the  {\it Schr\"odinger cat}
 structure of Eq. (\ref{E:loc}) can be extended to a
 whole interval of $T/U$ ratios, by replacing the Fock vectors $|{\cal
 N}_m\rangle$ with suitable su$(M)$ coherent states which are strictly
related to the localized ground state of the semiclassical approximation to
 Hamiltonian (\ref{E:BHH}).

\section{Time dependent variational principle}
\label{S:semicl}
As we recall above, the attractive Bose-Hubbard model (\ref{E:BHH}) has received significant attention also as the quantum counterpart \cite{A:Scott86} of the so-called discrete self-trapping (DST) or discrete non-linear Schr\"odinger (DNLS) equations \cite{A:Eilbeck85,A:Kevrekidis2,A:Eilbeck03}. These semiclassical equations are equivalently obtained by discretizing the Gross-Pitaevskii equations \cite{A:TrombL} or applying to Hamiltonian (\ref{E:BHH}) a time dependent variational principle (TDVP) based on Glauber's coherent states  \cite{A:Amico}. Here we adopt the same approach as Ref.~\cite{A:Amico}, using a macroscopic trial state of the form $|\tilde \Phi\rangle = e^{i {\cal S}} |\Phi\rangle $ where
\begin{equation}
\label{E:twf}
|\Phi\rangle \!=\! \frac{1}{\sqrt{{\cal N}! {\cal N}^{\cal N}}}\!\left(\sum_{m=1}^M \phi_m a_m^\dag \right)^{\cal N} \!\!\!\!|\Omega\rangle, \;\; \sum_{m=1}^M |\phi_m|^2 = {\cal N}, 
\end{equation}
instead of a product of Glauber's coherent states. The complex quantity $\phi_m$ appearing in Eq.~(\ref{E:twf}) is a dynamical variable describing the bosons  at lattice site $m$ through a macroscopic local phase and population, identified with $\arg(\phi_m)$ and $|\phi_m|^2$.
 The request that the trial state satisfies the Schr\"odinger equation on the average \cite{N:hbar}, $ \langle\tilde \Phi |i \partial_t -H|\tilde \Phi\rangle = 0 $, leads to the identification of $\dot {\cal S}$ with the effective Lagrangian for the dynamical variables $\phi_m$,  and 
\begin{eqnarray}
\label{E:scBH}
{\cal  H} &=& \langle \Phi |H|\Phi\rangle\nonumber\\ 
&=& \frac{{\cal N}-1}{\cal N} \frac{U}{2} \sum_{m=1}^M |\phi_m|^4  - T \sum_{m,m'=1}^M J_{m m'} \phi^*_m \phi_{m'} 
\end{eqnarray}
with the corresponding effective Hamiltonian \cite{A:Amico}. Minimizing $\cal H$ with respect to the $\phi_m$'s provides a semiclassical variational approximation to the ground state of the Bose-Hubbard Hamiltonian. Introducing the chemical potential $\mu$, i.e. a Lagrange multiplier allowing for the normalization constraint in Eq.~(\ref{E:twf}), the ensuing equations are 
\begin{equation}
\label{E:fpe}
 \frac{{\cal N}-1}{\cal N} U |\phi_m|^2 \phi_m  - T \sum_{m'} J_{m m'} \phi_{m'} =  \mu \phi_m
\end{equation}
We recall that Eqs. (\ref{E:fpe}) are actually the fixed point equations corresponding to the equations of motion for the $\phi_m$'s 
\begin{eqnarray}
\label{E:dst}
i \dot \phi_m \!\!&=&\!\! \{\phi_m^*,{\cal H}\}\nonumber\\ 
&=& \!\! \frac{{\cal N}-1}{\cal N} U |\phi_m|^2 \phi_m - T \sum_{m'} J_{m m'} \phi_{m'},
\end{eqnarray}
obtained via the canonical Poisson brackets
$\{\phi_m,\phi_{m'}^*\}=i\delta_{m\,m'}$.
Here by fixed points we mean the ``normal modes'' whose time evolution
is entirely contained in a simple phase factor $e^{-i \mu t}$.
Equations~(\ref{E:dst}) clearly belong to the DNLS class and, as we
show in the following section, they are able to capture many of the
features of the quantum ground state also in the extreme quantum regime of
large interactions and small populations. This is due to the extra
$({\cal N}-1)/{\cal N}$ factor appearing in their interaction term,
which is absent in the DNLS equations standardly adopted in the
semiclassical approach to the Bose-Hubbard model. The presence of this
extra factor is a direct consequence of  the macroscopic trial state, Eq.~(\ref{E:twf}), being an eigenstate of the total number
operator $N$. This is not true of the Glauber coherent state product
adopted as a trial function in standard TDVP, which conserves the
total number of bosons only on the average \cite{A:Amico}. Hence the
resulting dynamical equations, which are perfectly equivalent to
standard DNLS equations, do not feature the extra $({\cal N}-1)/{\cal
  N}$ factor. The effectiveness of the description provided by the
macroscopic state (\ref{E:twf}) can be explained by noticing
that it is actually a  coherent state itself, of su$(M)$. 
Note that $\cal N$ is the algebra representation index of the
present two-boson realization of su$(M)$ as confirmed by the fact
that $N$ commutes with the algebra generators $a^+_m a_{m'}$
for any $1 \leq m, m' \leq M$.  Hence the
conservation of the  boson population is already incorporated in this
coherent state picture . This is illustrated in Appendix \ref{S:suM}

\begin{figure}
\begin{center}
\includegraphics[width=8.5cm]{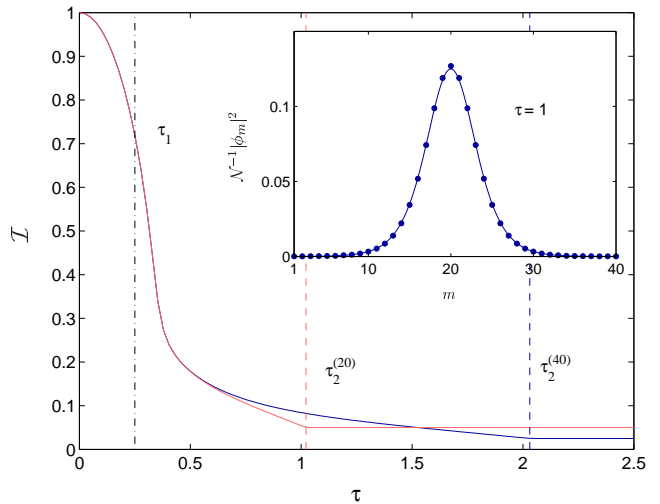}
\caption{\label{F:regimes} (Color online) Normalized interaction energy, Eq.~(\ref{E:int_en}), for two different lattice sizes,  $M=20$ (light solid line) and $M=40$ (dark solid line). The delocalization transition clearly occurs at $\tau = \tau_2^{(M)}$. Furthermore, the plotted quantity exhibits a size independent behavior in the small $\tau$'s regime. The crossover at $\tau=\tau_1$ corresponds to a change in the convexity plotted function. This is related to the ultraviolet cutoff due to the discreteness of the lattice. Note indeed that in the continuum limit the corresponding quantity would diverge  as $\tau^{-1}$ for $\tau \to 0$. The inset shows a comparison between the localized ground state of $\cal H$ (bullets) on a lattice of size $M=40$ and the corresponding solution in the continuum limit (solid line).}
\end{center}
\end{figure}

As it is well known,  the nonlinear terms can break the translational
symmetry characterizing Hamiltonian (\ref{E:scBH}) with the choice in Eq.~(\ref{E:homJ}). Indeed, if the interaction $U$ is sufficiently strong, 
the  ground state of the system is arbitrarily localized at one of the $M$ equivalent lattice sites. This ground state is a fixed point of Eqs. (\ref{E:dst}), and can be easily determined numerically by treating Eqs.~(\ref{E:fpe})
as a set of self-consistent linear equations \cite{A:Kevrekidis}. 
More in general, depending on the value of the $T/U$ ratio, it is
possible to recognize three regimes for the semiclassical ground
state. For large ratios, the ground state is superfluid, the boson
population being evenly spread over the entire lattice. For
sufficiently low values of the $T/U$ ratio this delocalized superfluid
solution becomes unstable, and the ground state of the system becomes
soliton-like, featuring a localization peak breaking the translational symmetry of Hamiltonian (\ref{E:scBH}).
The critical point marking the boundary between the {\it superfluid} and
{\it soliton regime}  is $\tau = \tau_2^{(M)} =[ 2 M \sin^2 (\pi/M)]^{-1}$, where we recall that we define $\tau = T/|U| ({\cal N}-1)$ \cite{A:Jack,A:Smerzi2,N:trans}. 
As $\tau$ is further decreased, the localization peak --- which
initially involves a significant number of lattice sites --- becomes
sharper and sharper, until the entire boson population is basically
concentrated at a single lattice site. This situation is the is the 
semiclassical, symmetry-breaking counterpart of the Schr\"odinger-cat
state, where the boson localization is symmetrically distributed over
all of the lattice sites. As we discuss in the sequel, it is possible
to recognize a threshold, or rather a crossover, between this
{\it Schr\"odinger-cat regime} and the soliton regime  where the
localization peak involves a large portion of the lattice. 
Note indeed that in the latter situation the periodic boundary conditions and the discretization of the lattice produce a small effect, and the ground state is well described by the solution to the nonlinear Schr\"odinger equations on a 1D infinite domain, $\phi_m = e^{i \varphi} \sqrt{{\cal N}/2\Delta }\, {\rm sech}(m/\Delta)$, where $\Delta = 4 T/|U| ({\cal N}-1) = 4 \tau$ is the characteristic width of the population
peak. The inset of Fig.~\ref{F:regimes} shows that this is actually a
satisfactory approximation.  Hence, the crossover between the
Schr\"odinger cat and soliton regime can be estimated by requiring that this characteristic width equals the lattice spacing, i.e. $\Delta \approx 1$, which results into $\tau = \tau_1 \approx 1/4$. Note that this is equivalently obtained by equalling the energies of the system in the two extreme situations $T \ll |U|$ and $T \gg |U|$. Indeed in the former case the ground state is completely localized and its energy is ${\cal N}({\cal N}-1)U/2$, while in the latter case the ground state is delocalized and its energy is $-2 T {\cal N}$. We emphasize that exactly the same estimate is obtained considering the energy of the quantum ground state.
Conversely, the delocalization transition at $\tau=\tau_2^{(M)}$,
marking the boundary between the soliton and superfluid regime,
corresponds to the peak width being of the same order of the lattice
size, $\Delta \approx M$. Note indeed that $\tau_2^{(M)}$ is linear in
$M$ for sufficiently large lattices \cite{N:Kanamoto2}.
The regimes discussed above are quite evidently signalled by the behavior of the (normalized) interaction energy
\begin{equation}
\label{E:int_en}
{\cal I} = \frac{\sum_{m=1}^M |\phi_m|^4}{{\cal N}^2} = \frac{M}{{\cal N}^2} \langle n_m^2 -n_m\rangle
\end{equation}
as a function of $\tau$, where $\langle \cdot \rangle$ denotes expectation value on the trial state (\ref{E:twf}) equipped with the ground state solution to the fixed point equations (\ref{E:fpe}). 
This is shown in Fig.~\ref{F:regimes}, where two lattice size are considered, $M=20$ and $M=40$. 
Recalling that $\langle n_m \rangle = {\cal N}/M$, it is easy to see that the quantity in Eq. (\ref{E:int_en}) is also related to the fluctuation of the local density of bosons.

\section{Semiclassical vs. Quantum Picture}
\label{S:results}
In this section we compare the features of the semiclassical picture introduced in the preceding Section to the corresponding features of the quantum model, Eq. (\ref{E:BHH}).
First, we direct our attention to the energy and  structure of the ground state. Then, we discuss the relevance of the semiclassical thresholds in the quantum picture. We consider lattice comprising up to $M=20$ sites, and populations consisting of up to ${\cal N} = 10$ bosons. All of the quantum observables are obtained evaluating the eigenstates of Hamiltonian (\ref{E:BHH}) and the relevant energies by means of Lanczos-like algorithms. The translational symmetry of Hamiltonian (\ref{E:BHH}) is conveniently taken into account in order to reduce significantly the computational demand.
\subsection{Ground-state energy and structure}
\begin{figure}
\begin{center}
\includegraphics[width=8.5cm]{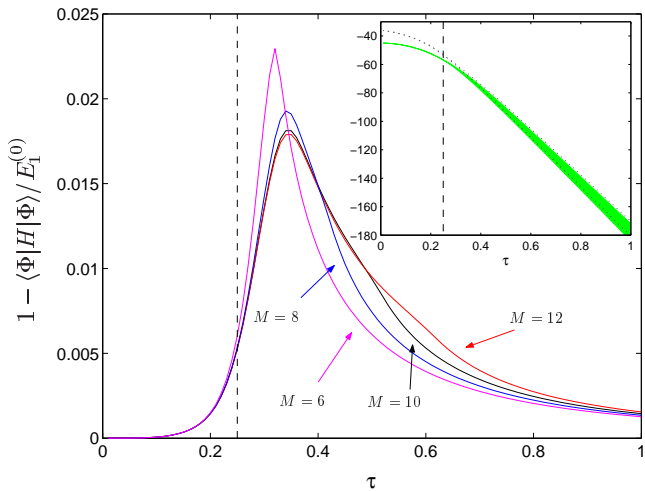}
\caption{\label{F:deltaE} (Color online) Relative energy difference between the exact quantum ground state and the relevant semiclassical approximation. The curves refer to boson population ${\cal N} = 10$ and lattice sizes $M = 6,8,10,12$. The inset shows the energy range spanned by the ground states $E_1^{(k)}$ (shaded region) and the first excited states  $E_2^{(k)}$ (region between the dotted lines) of the blocks relevant to different values of the quantum number $k$ associated to the translational invariance of $H$ [see Eq. (\ref{E:Hspec})]. These result refer to $M =12$ and $N=10$. The vertical dashed lines signal the crossover at $\tau=\tau_1$. }
\end{center}
\end{figure}

\begin{figure}
\begin{center}
\includegraphics[width=8.5cm]{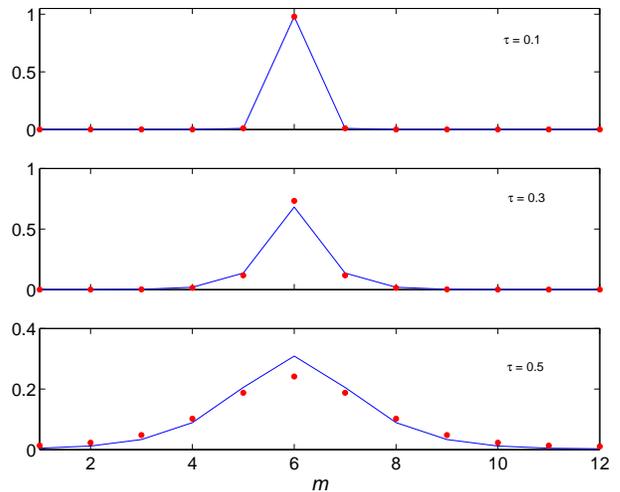}
\caption{\label{F:loc_st} (Color online) Normalized boson density $\langle\Phi_j|n_m|\Phi_j\rangle$  as provided by the quantum localized state (solid lines), compared to the corresponding semiclassical quantity $|\phi_m|^2$ (bullets) for $N=10$ bosons on $M=12$ lattice sites.   }
\end{center}
\end{figure}

As it can be seen in Fig.~\ref{F:deltaE}, the ground state energy of the quantum model, Eq.~(\ref{E:BHH}) is in general remarkably well described by the relevant semiclassical result. 
Indeed the maximum in the relative error settles at about 2\% as the
lattice size increases. Furthermore, the agreement between the quantum
and the classical result is extremely satisfactory both in the large
$\tau$ and in the small $\tau$ region. The agreement in the former
region is expected, since in the noninteracting limit the quantum
ground state of the system is exactly given by Eq.~(\ref{E:twf}) with
$\phi_m = \sqrt{{\cal N}/M}$, which is nothing but the superfluid
semiclassical ground state taking over after the delocalization
transition at $\tau_2^{(M)}$. The agreement in the strong interaction
limit is more surprising, and it requires further discussion. Indeed, in this region the semiclassical and the quantum ground states are  significantly different, the former being  strongly localized at one lattice site, while the latter being  always completely delocalized. As we recall in Sec.~\ref{S:Quantum}, the eigenvalues of the quantum Hamiltonian (\ref{E:BHH}) are labeled by a quantum number $k$ ensuing from translational invariance, Eq. (\ref{E:Hspec}).
The inset of Fig.~\ref{F:deltaE} clearly shows that, in the small $\tau$ region the actual ground state, $|\psi_1^{(0)}\rangle$, is almost degenerate with the ground states of all of the remaining Hamiltonian blocks, $|\psi_1^{(k)}\rangle$. 
Hence, in this region, the actual ground state of the system 
is satisfactorily described by any superposition of the states $|\psi_1^{(k)}\rangle$. Of course this is not an eigenstate of $H$, but the relevant dephasing time is extremely long, being inversely proportional to the energy spreading of the involved eigenstates. In particular, 
one can choose a superposition of the form \cite{A:Scott94}
\begin{equation}
\label{E:coh_sup}
|\Phi_j \rangle = \sum_{k=1}^{M} e^{-i \frac{2 \pi}{M} j\, k} \gamma_k |\psi_1^{(k)}\rangle
\end{equation}
which features a population peak localized at lattice site $j$. We emphasize that the phase factors $\gamma_k$ are not arbitrary, but have to be chosen to make sure that the  $|\psi_1^{(k)}\rangle$'s are phase coherent. Recall indeed that a generic eigenstate is defined up to an arbitrary phase factor. The superposition in Eq.~(\ref{E:coh_sup}) features the desired localization properties 
only if the $\gamma_k$'s compensate for possible phase differences among the $|\psi_1^{(k)}\rangle$'s \cite{N:phase2}.
 As it is shown in Fig.~\ref{F:loc_st}, the boson density
 $\langle\Phi_j|n_m|\Phi_j\rangle$  is in remarkable agreement with
 the corresponding semiclassical quantity, $|\phi_m|^2$ for
 sufficiently small values of $\tau$. At larger values the agreement
 is lost since the energy spreading of the eigenstates involved in
 Eq.~(\ref{E:coh_sup}) becomes significant, and the superposition
 cannot be confused with the ground state any more. Furthermore, as it
 is shown in the inset of Fig.~\ref{F:deltaE},  the energies of the
 $|\psi_2^{(k)}\rangle$'s become comparable to those of the
 $|\psi_1^{(k)}\rangle$ 's. 

By resorting to perturbation theory, it is possible to show that the agreement between the trial state~(\ref{E:twf}) and the localized state $|\Phi_j\rangle$, Eq. (\ref{E:coh_sup}), goes beyond the local density of bosons. Indeed, at the first order in the perturbative parameter $\tau$, they are actually the same state. This is discussed in  Appendix \ref{S:pert}. 
Notice that  the above discussion entails that, in the strong coupling, Schroedinger-cat regime, the actual ground state of the quantum Hamiltonian (\ref{E:BHH}) is a coherent superposition of localized states of the form (\ref{E:twf}), where the $\phi_m$'s are obtained solving the corresponding semiclassical problem (\ref{E:fpe}).
Once again, we emphasize that the extra $({\cal N}-1)/{\cal N}$ in the
DNLS equations (\ref{E:dst}), stemming from the choice of a
number-conserving trial state, Eq.~(\ref{E:twf}), plays a
fundamental role in ensuring the remarkable agreement between the
quantum and the semiclassical picture, Eqs. (\ref{E:BHH}) and
(\ref{E:scBH}), respectively.

\subsection{Quantum signature of the semiclassical regimes}
Let us now discuss the relevance of the semiclassical thresholds $\tau_1$ and $\tau_2^{(M)}$ in the quantum picture of the system. This issue has been considered to some extent in Ref.~\cite{A:Jack}, where only the delocalization transition occurring at $\tau = \tau_2^{(M)}$ is taken into account. The authors report that this critical point is signalled by the behavior of some quantum observables, such as the one-particle and ${\cal N}$-particle correlation functions,
\begin{equation} 
\label{E:corfun}
C_1 =\frac{M}{\cal N} \langle a^\dag_m a_{m+1}\rangle,\quad C_{\cal N} = \frac{M}{{\cal N}!} \langle \left(a^\dag_m\right)^{\cal N} a_{m+1}^{\cal N}\rangle,
\end{equation}
where $\langle \cdot \rangle$ denotes expectation value over the ground state. More precisely, they observe that, with increasing interaction,  $C_1$ features a sudden drop around the semiclassical delocalization threshold $\tau_2^{(M)}$, while $C_{\cal N}$ features a surge roughly at the same point. Furthermore they report that this threshold is also signalled by the number distribution, i.e. the probability $\rho_\nu$ that the boson population at a given lattice site equals $\nu$. More precisely, they report that this distribution becomes double peaked for $\tau<\tau_2^{(M)}$. 

Here we show that the above quantum observables do not signal the delocalization threshold, but rather the crossover between the soliton and Schr\"odinger-cat regime occurring around $\tau=\tau_1$. First of all, we observe that the variation of these observables is not so sudden to identify a precise transition point. They rather signal a crossover interval. Furthermore, in the case considered in Ref.~\cite{A:Jack}, $M=6$, $N=10$, the two semiclassical thresholds $\tau_1 =1/4$ and $\tau_2^{(6)} = 1/3$ are relatively close, and both compatible with such crossover interval. However, as we illustrate in Section~\ref{S:semicl}, $\tau_1=1/4$ is substantially size-independent, while $\tau_2^{(M)}$ clearly depends on the lattice size. A study of the scaling properties of the above quantum observables shows that the features considered in Ref.~\cite{A:Jack} do not depend on the lattice size, and occur in the region around $\tau_1$ \cite{N:Jack}.
\begin{figure}
\begin{center}
\includegraphics[width=8.5cm]{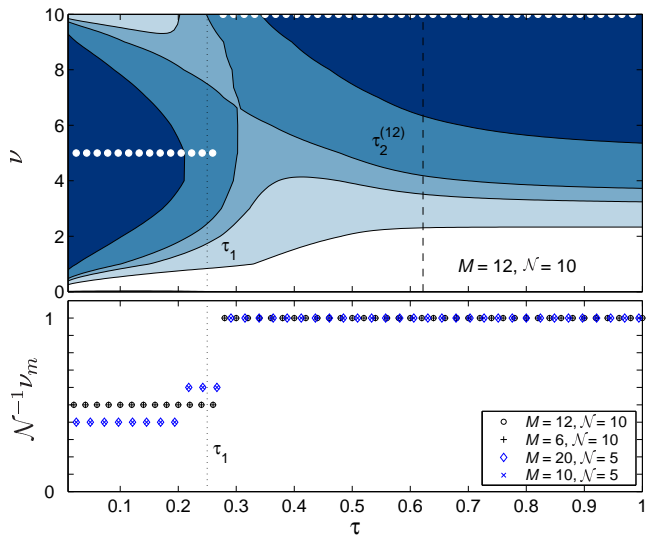}
\caption{\label{F:rho_n} (Color online) Top: contour plot of the number density $\rho_\nu(\tau)$ for ${\cal N}=10$ bosons on a lattice of size $M=12$. The darker the shade, the lower the density. The white bullets signal the value $\nu_{\rm m}(\tau)$ at which the density is minimum. The vertical  lines signal the semiclassical thresholds. Bottom: plot of $\nu_{\rm m}(\tau)$ for several choices of $\cal N$ and $M$.  }
\end{center}
\end{figure}

Let us first of all consider the number distribution $\rho_{\nu}$, i.e. the quantity such that $\langle n_m \rangle = \sum_{\nu = 0}^{\cal N} \nu\, \rho_{\nu} = {\cal N}/M$ and, in general,  $\langle n_m^p \rangle = \sum_{\nu = 0}^{\cal N} \nu^p\, \rho_{\nu}$. It is clear that in the strong interaction limit this distribution features two peaks at $\nu=0$ and $\nu = \cal N$, while it is substantially vanishing for all of the remaining numbers \cite{A:Jack}. Recall indeed that in this limit, , i.e. deep in the Schr\"odinger-cat regime, the ground state has the form in Eq.~(\ref{E:loc}) \cite{A:Jack,A:Eilbeck85}. In this regime the number fluctuations are very large. Actually, as we discuss in Sec. \ref{S:semicl}, this situation is the quantum counterpart of the semiclassical situation where the boson population is almost entirely concentrated at one of the lattice sites. Hence, the ground state cannot be influenced by the boundary conditions or by the lattice size, and this regime is expectedly entered at $\tau \approx \tau_1$. This is actually confirmed by Fig.~\ref{F:rho_n}, which illustrates some key features of the number distribution. As it can be seen in the contour plot of  in the top panel, for small $\tau$ the number distribution $\rho_\nu(\tau)$ features two maxima at the extreme values  $\nu =0$ and $\nu = {\cal N}$, while its minimum is at $\nu = {\cal N}/2$. As soon as $\tau$ becomes larger than $\tau_1$ the relative maximum at $\nu = {\cal N}$ disappears, and the minimum suddenly moves to this extreme value. The location of such minimum, $\nu_{\rm m}$, is signalled by the white bullets on the contour plot. Note that no special feature can be recognized in correspondence of the delocalization threshold, signalled by the vertical dashed line. The bottom panel of Fig.~\ref{F:rho_n} reports the (normalized) location $\nu_{\rm m}(\tau)/\cal N$ of the minimum of $\rho_\nu(\tau)$ for several choices of the lattice size $M$ and of the boson population $\cal N$. In all cases the sudden displacement of the minimum signals that the number distribution becomes single peaked at the semiclassical crossover $\tau_1$, i.e. at the crossover between the Schr\"odinger-cat and soliton regime.
\begin{figure}
\begin{center}
\includegraphics[width=8.5cm]{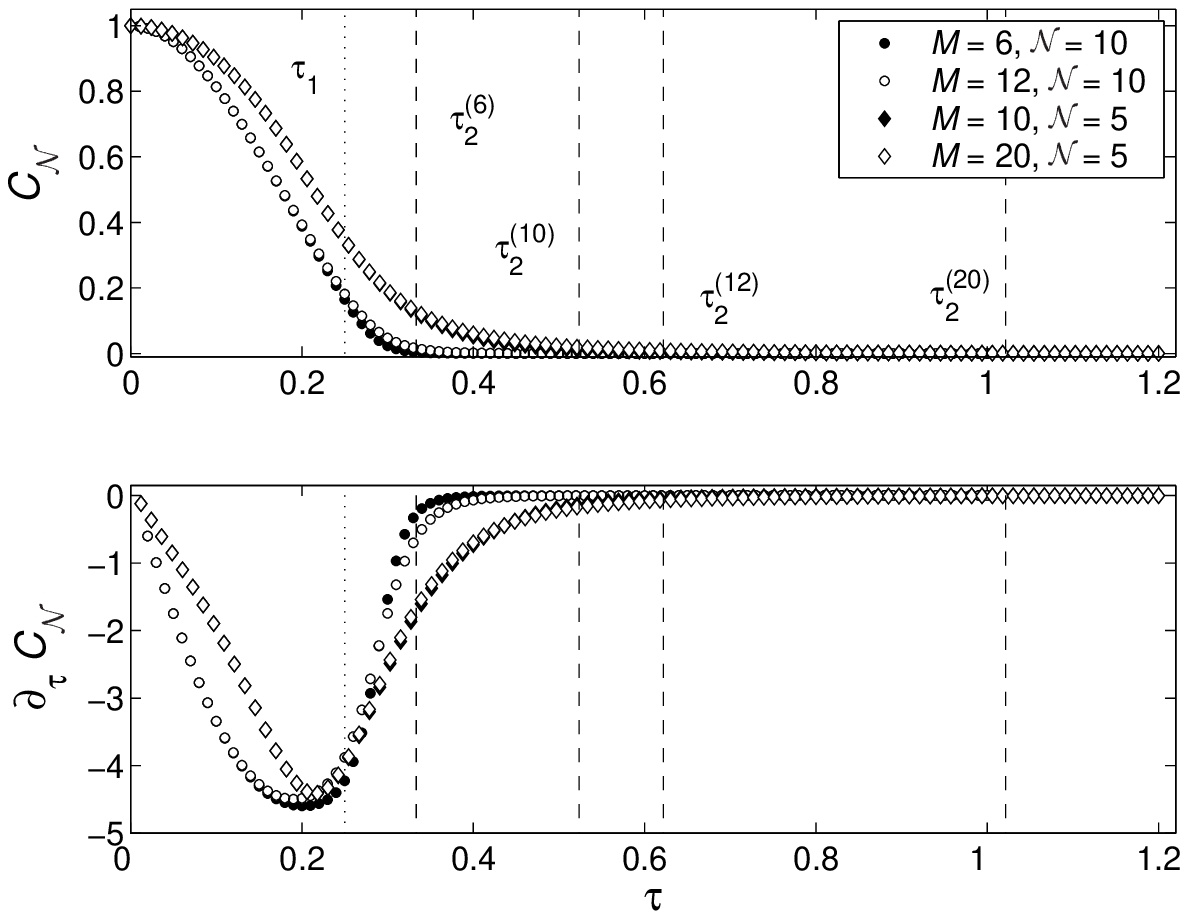}
\caption{\label{F:cN} Behavior of the  ${\cal N}$-particle correlation function  $C_{\cal N}$ (top) and of its first derivative (bottom) with respect to $\tau$. The dotted vertical lines signal the semiclassical crossover at $\tau_1$. The dashed vertical lines signal the semiclassical delocalization thresholds relevant to the lattice sites considered $M = 6, 10, 12, 20$.   }
\end{center}
\end{figure}
Basically the same discussion as above applies to the ${\cal N}$-particle correlation function  $C_{\cal N}$ defined in Eq.~(\ref{E:corfun}). It is indeed clear that it must exhibit a significant surge when the Schr\"odinger-cat regime is entered and the number distribution becomes double peaked, i.e. around $\tau = \tau_1$. Indeed, for this correlation to be non-vanishing it is necessary that the probability of finding all of the $\cal N$ bosons at the same lattice site is significant. This is confirmed by Fig.~\ref{F:cN}, reporting the behavior of $C_{\cal N}(\tau)$ and of its derivative with respect to $\tau$ for several choices of the boson population and of the lattice size. No particular feature occurs at the $M$ dependent delocalization thresholds, signalled by the vertical dashed lines. Actually, a dependence on the lattice size $M$ is hardly recognizable,  the curves referring to the same number of bosons ${\cal N}$ being almost perfectly coincident.  Conversely, the location where the plotted quantity becomes significant and its concavity changes are satisfactorily estimated by the semiclassical size-independent crossover at $\tau=\tau_1$. 
\begin{figure}
\begin{center}
\includegraphics[width=8.5cm]{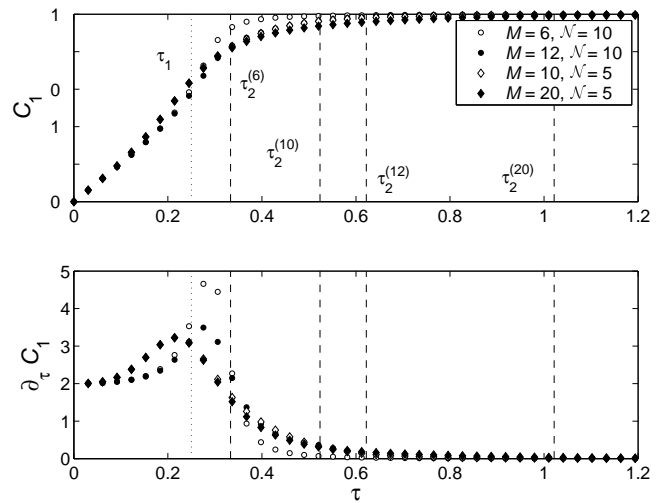}
\caption{\label{F:c1} Behavior of the  single-particle correlation function  $C_1$ (top) and of its first derivative (bottom) with respect to $\tau$. The dotted vertical lines signal the semiclassical crossover at $\tau_1$. The dashed vertical lines signal the semiclassical delocalization thresholds relevant to the lattice sites considered $M = 6, 10, 12, 20$. }
\end{center}
\end{figure}
Figure~\ref{F:c1} shows that also the substantial drop in the single-particle correlation function $C_1$ considered in Ref.~\cite{A:Jack} actually occurs around $\tau = \tau_1$. This is confirmed by the behavior of the derivative of this quantity, shown in the bottom panel of the same figure.
Again, the plotted quantity is not sensitive to the size-dependent delocalization threshold $\tau_2^{(M)}$, essentially due to the fact that it is a local quantity, as it can be seen from its definition, Eq.~(\ref{E:corfun}).

\begin{figure}
\begin{center}
\includegraphics[width=8.5cm]{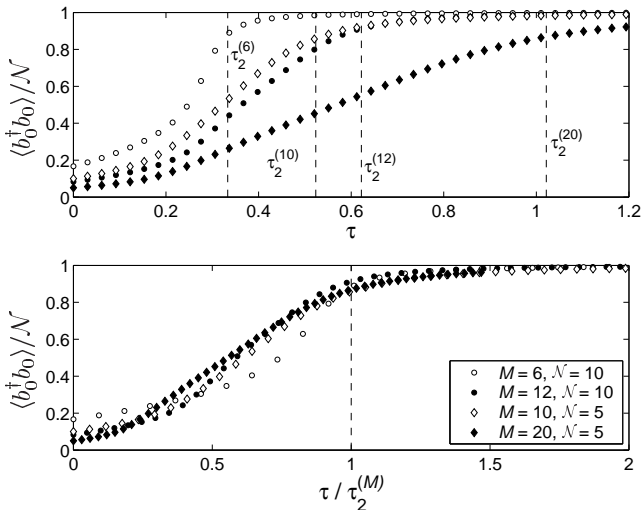}
\caption{\label{F:nk0} Behavior of the  occupation of the lowest reciprocal mode  $b_0^\dag b_0$ with respect to $\tau$ (top) and with respect to $\tau/\tau_2^{(M)}$. The dashed vertical lines signal the semiclassical delocalization thresholds relevant to the lattice sites considered $M = 6, 10, 12, 20$. The data collapse in the bottom panel shows that $b_0^\dag b_0$ actually signals the semiclassical delocalization threshold at $\tau_2^{(M)}$. }
\end{center}
\end{figure}

The size-dependent delocalization threshold between the soliton and superfluid regime is signalled by a typically nonlocal quantity such as the occupation number $b_0^\dag b_0$  of the lowest reciprocal mode, where $b_k$ is defined in Eq.~(\ref{E:bk}). This is clearly evident in Fig.~\ref{F:nk0}, where $\langle b_0^\dag b_0 \rangle/{\cal N}$ is plotted with respect to both $\tau$ (top panel) and $\tau/\tau_2^{(M)}$ (bottom panel), for several lattice sizes and boson populations. Unlike the quantities in Figs.~\ref{F:rho_n}-\ref{F:c1}, $\langle b_0^\dag b_0 \rangle/{\cal N}$ exhibits a clear dependence on the lattice size, when plotted against $\tau$. Conversely, when the abscissae of the plot are rescaled so that all of the size-dependent delocalization thresholds occur at the same point (signalled by the vertical dashed line), a very satisfactory data collapse becomes evident. This means that the drop in the occupation of the lowest reciprocal mode is related to the delocalization transition observed in the semiclassical model.
\section{Outlook}
In this paper we consider the Bose-Hubbard Hamiltonian (\ref{E:BHH}) describing ultracold attractive bosons trapped into the {\it necklace-like} lattice  whose realization in terms of optical potentials is described in Ref. \cite{A:Amico05}. In particular, we study the properties of the ground-state  of such extremely symmetric quantum Hamiltonian, which in the strong interaction limit is known to exhibit the features of a {\it Schr\"odinger-cat} state.
We introduce a variational approximation to this ground state  based
on a su$(M)$ coherent state, Eq. (\ref{E:twf}). 
The equations governing the dynamics of the labels of this coherent
state belong to the class of DNLS equations, yet they feature an
unusual correction term ensuing from the total number
conservation. Owing to this correction term, the ground-state of this
semiclassical approximation provides a remarkably satisfactory
description of its quantum counterpart not only --- as expected --- in the small-interaction superfluid regime, but also in the deeply quantum 
Schr\"oedinger-cat regime, i.e. for strong attractive interactions.
In particular we show that in this regime the delocalized quantum ground state of the system can be seen as a coherent superposition of the localized ground states of the variational approach. Furthermore, we show that the three regimes clearly recognizable in the semiclassical picture are signalled by suitably chosen quantum observables. The latter result is compared with similar findings reported in Ref. \cite{A:Jack}.
In the intermediate regime  the description provided by the above semiclassical picture  is not as satisfactory as in the two outermost regimes.  The structure of the quantum ground state in this parameter range is different, and probably much more complex, than that of the variational trial state (\ref{E:twf}), and requires further investigations.


Before concluding this paper we briefly consider some experimental issues. First of all, we observe that attaining the deeply quantum Schr\"odinger-cat regime might prove very challenging from the experimental point of view. Indeed, this would be most likely obtained by starting from the ground state of the (almost) noninteracting superfluid regime, which is relatively easy to realize. By making use of Feshbach resonances and/or by increasing the strength of the optical lattice, the ratio $T/|U|$ can be subsequently decreased down to  suitably small values, so that the system is driven into the Schr\"odinger-cat regime.  Note that, independent of the value of such ratio, the ground state of the system has always the same symmetry, i.e. it always belongs to the same eigenspace of the displacement operator defined in Eq. (\ref{E:disp}). This means that, at least in principle, only the relevant block of Hamiltonian (\ref{E:BHH}) is involved in the process. Actually, this is true provided that the system is perfectly isolated from the environment. Under this rather ideal assumption, the achievement of the Schr\"odinger-cat ground state  basically depends on the adiabaticity of the parameter variation. However, the interaction with the environment much likely introduces a coupling between the ideally uncoupled blocks of Hamiltonian (\ref{E:BHH}), which has dramatic effects in the strong interaction regime. Indeed in this situation the actual ground state  is almost degenerate with the lowest states of the blocks relevant to $k = 2,\ldots, M$, and an arbitrarily small perturbation can induce a transition to one of these  excited states.

Another issue worth mentioning is the actual possibility of making measurements on a system such as the one considered here. As we recall in Section \ref{S:intro}, there is an upper bound to the number of attractive bosons $\cal N$ that can be loaded into the optical lattice. Indeed, the maximum attainable boson density $\langle n_m\rangle = {\cal N}$ must be sufficiently low to rule out the complex phenomena described in Ref. \cite{N:collapse}. On a one-dimensional system like the one considered here, the maximum boson population compatible with such constraint can be too small to allow measurements, , such as e.g. absorption imaging. This difficulty can be possibly circumvented by considering a stack of coaxial {\it necklace} optical lattices isolated from each other, in place of a single necklace \cite{A:Amico05}. Assuming that these lattices can be loaded with the same number of bosons, they are expected to behave like copies of the system discussed in this paper. On the other hand, the larger overall boson population should allow for effective measurements.

\acknowledgments
The authors wish to thank F. Minardi, P. Maioli and P.G. Kevrekidis for stimulating discussions. One of the authors (P.B.) acknowledges a grant from the {\it Lagrange Project} - CRT Foundation.

\appendix
\section{su($M$) coherent states }
\label{S:suM}
In this section we give a sketch of the proof that the variational trial state introduced in Eq.~(\ref{E:twf}) is actually a su$(M)$ coherent state, where $M$ is the number of sites composing the lattice.
According to the standard theory  \cite{A:Zhang,B:Perelomov}, the
coherent states of a given group $G$ are represented, up to a phase
factor, as $|{\rm cs}\rangle = g |{\rm ext}\rangle$, where $g \in G$ and $|{\rm
  ext}\rangle$ is a so-called {\it extremal vector}. The latter is
identified by determining the maximal isotropy sub algebra ${\cal
  B}\in{\cal G}^C$, where ${\cal G}^C$ is the complexified version of
the algebra $\cal G$ associated to the group $G$.

In our particular case, $\cal G$ is the su$(M)$ algebra represented in
terms of the $M$ lattice boson operators $a_m$. Hence ${\cal G} = \{
h_s, a_m^\dag a_{m'}: \, m \ne m'; \,  m, m'\, = 1,2,\ldots, M; \, s = 1,2,\ldots, M-1 \}$
and a generic element of the SU$(M)$ group is given by 
\[
g =e^{  i \left [ \sum_s x_s h_s + 
\sum_{m< m'} (\sigma_{m\,m'} a_m^\dag a_{m'} +\sigma^*_{m\,m'} a_m^\dag a_{m'} ) \right ]
}
\]
where the Cartan operators $h_s$ are suitable linear combinations of
the on-site number operators $n_m=a^\dag_m a_m$, $x_m \in {\mathbb R}$
and $\sigma_{m\,m'} \in {\mathbb C}$. The relevant maximal isotropy sub algebra
$\cal B$ is formed by (complex valued) linear combinations of all of
the generators of $\cal G$, except the $M-1$ operators of the form 
$a^\dag_m a_{\bar  m}$, where $\bar m$ corresponds to one of the lattice
sites, and $m \ne \bar m$. The extremal state $| {\rm ext} \rangle$, is defined
as the state such that $b |{\rm ext} \rangle = \lambda_b |{\rm ext}\rangle$ for any 
$b \in {\cal G}\cap {\cal B}$, with $\lambda_b \in {\mathbb C}$
\cite{B:Perelomov}. This  yields  $|\rm ext\rangle = |{\cal  N}_{\bar m}\rangle$, where $|{\cal   N}_{\bar m}\rangle$ 
is defined in Eq.~(\ref{E:loc}). Of course, all of
the $M$ possible choices for $\bar m$, and hence for the extremal
state, are equivalent. For definiteness, here we set $\bar m = 1$.

It is possible to show that for any $g \in G$ there exists an unique decomposition 
$g=g_z e^b$, where  $b \in {\cal G}\cap {\cal B}$ and the so-called {\it displacement operator} has the form
\begin{equation}
g_z = \exp\left[\sum_{m =2 }^{M} z_m a^\dag_m a_{1}+ z^*_m  a^\dag_{1} a_{m} \right] .
\label{E:CS}
\end{equation}
Hence,  up to an irrelevant phase factor
arising from $e^b$, one gets  $| {\rm cs} \rangle = g | {\rm ext} \rangle = g_z | {\cal N}_1 \rangle$ 

On the other hand, the variational trial state introduced in Eq.~(\ref{E:twf}) can be rewritten as
\begin{equation}
|\Phi \rangle = \frac{(A^\dag)^{\cal N}}{\sqrt{{\cal N}!}} |\Omega\rangle, \quad A = \frac{1}{\sqrt{\cal N}}\sum_{m=1}^M \phi_m a_m
\label{E:twf2}
\end{equation} 
Observing that $A^\dag = B a_1^\dag B^\dag$, and hence that $(A^\dag)^{\cal N} = B (a_1^\dag)^{\cal N} B^\dag$, where
\begin{eqnarray}
&B = e^{i \sum^M_{m=1} \varphi_m n_m } 
e^{i\sum^M_{m=2} \theta_m (a_m^\dag a_1 + a_1^\dag a_m) },& \nonumber\\
&\frac{\phi_1}{\sqrt{\cal N}}= e^{i\varphi_1} \cos \theta, \;\; \frac{\phi_k}{\sqrt{\cal N}}= i e^{i\varphi_k} \theta_k \frac{\sin \theta}{ \theta},\;\;
\theta^2 = \sum^M_{m=2} \theta^2_m, \nonumber
\end{eqnarray}
one gets
\begin{equation}
|\Phi \rangle =  B \frac{(a_1^\dag)^{\cal N} }{\sqrt{{\cal N}!}}\,|\Omega\rangle =  B |{\cal N}_1 \rangle,
\label{E:twf3}
\end{equation} 
where we made use of the fact that $B^\dag |\Omega\rangle = |\Omega\rangle$.
After some further algebraic manipulations, 
state (\ref{E:twf3}) reduces to (up to an irrelevant phase factor)
$$
|\Phi \rangle = e^{i\varphi_1 {\cal N}}
e^{\sum^M_{m=2} (\zeta_m a_m^\dag a_1 + \zeta^*_m a_1^\dag a_m)} \, |{\cal N}_1\rangle \, ,
$$
where $\zeta_m = \theta_m e^{i\varphi_m} \in {\bf C}$.
The comparison of the latter expression with Eq. (\ref{E:CS})
fully discloses the SU($M$)-coherent-state structure of the variational trial
state $|\Phi \rangle$ defined in Eq. (\ref{E:twf}).

It is important to notice that, while 
$|\Phi \rangle$ defined as 
in Eq. (\ref{E:twf2}) is operationally
useful to derive the semiclassical picture relevant to
Eqs. (\ref{E:scBH}) and (\ref{E:dst}), 
the hidden, equivalent form of $|\Phi \rangle$ as a SU$(M)$ coherent-state (CS)
indeed is responsible for such a semiclassical scenario.

In general, CS are known to provide a formulation of
quantum-mechanical problems \cite{A:Zhang,B:Perelomov}
particularly close to the corresponding classical Hamiltonian picture.
Let us consider   a classical Hamiltonian $H_{\rm C}$ for $2 M$ degrees of freedom and  the corresponding quantum model $H_{\rm Q}$ obtained via canonical
quantization. The dynamics of the quantum model can be investigated enacting
a time-dependent variational principle (TDVP) based on a coherent state  $|\Xi \rangle$, with $\Xi = (\xi_1, \ldots, \xi_M) \in {\mathbb C}^M$. 
This allows one to recast the Schr\"odinger equation 
for $H_{\rm Q}$ into a classical problem whose Hamiltonian is 
${\cal H} (\Xi) = \langle \Xi |H_{\rm Q}|\Xi \rangle$.
The corresponding Hamiltonian equations describe the evolution of the CS labels $\xi_m$, which thus play the role of dynamical parameters. Simple analytic formulas show that $\xi_m$'s can be rewritten 
in terms of the canonical variables appearing in $H_{\rm C}$ (see, e. g., Ref. \cite{A:Amico}). More in general, the fact that $|\Xi \rangle$ is a coherent state ensures that the domain where $\Xi$ ranges (CS manifold) coincides with the phase space where the dynamics of $H_{\rm C}$ is represented.

The CS variational method exhibits two further
noticeable features. First,
the classical Poisson brackets characterizing the $\Xi$ dynamics preserve 
the algebraic structure of the operators constituting $H_{\rm Q}$ and
generating the group $G$. Also, the coherent-state labels $\xi_m$, beyond their
canonical role in the ${\cal H}$ dynamics, are
the expectation values of Hamiltonian operators 
that embody information on microscopic interaction processes.
For example, in the specific case under investigation,
$\langle \Phi |a^+_m a_{k} |\Phi \rangle= \phi^*_m \phi_k$,
$\langle \Phi | n_m |\Phi \rangle= |\phi_m|^2$ in Hamiltonian (\ref{E:scBH})
are manifestly related to
$a^+_m a_{k}$ and $n_m$ in Hamiltonian (\ref{E:BHH}).
Notice that Hamiltonian (\ref{E:scBH}) matches exactly the classical
version of Eq. (\ref{E:BHH}) as soon as ${\cal N}/({\cal N}-1) \simeq 1$.
An interesting results in the present work is the profound similarity 
between the (purely quantum) properties exhibited by the ground state of 
Hamiltonian (\ref{E:BHH}) and the properties of its semiclassical
minimum-energy configuration derived from state (\ref{E:twf2}) in the Schr\"odinger-cat regime.

\section{Perturbative comparison of the quantum and semiclassical ground states }
\label{S:pert}
In this section we show that in the strong interaction limit the superposition defined in Eq.~(\ref{E:coh_sup}) coincides with the coherent state (\ref{E:twf})
where the $\phi_m$'s correspond to the semiclassical  solution to Eqs.~(\ref{E:fpe}) localized at lattice site $j$. We do this resorting to perturbative theory, i.e. by assuming that
\begin{equation}
\label{E:H0}
H_0= \frac{U}{2} \sum_{m=1}^M n_m (n_m-1) 
\end{equation}
and
\begin{equation}
\label{E:pert}
 V = - T \sum_{m,m'=1}^M J_{m m'} a^\dag_m a_{m'}
\end{equation}
are the unperturbed and perturbative terms of the quantum Hamiltonian (\ref{E:BHH}), respectively. 
At the $0^{\rm th}$ order in $\tau = T/U ({\cal N}-1)$ the quantum superposition $|\Phi_j\rangle$ in Eq.~(\ref{E:coh_sup}) coincides with the Fock state $|{\cal N}_j\rangle$  defined in Eq.~(\ref{E:loc}). The latter  belongs to the $M$-fold degenerate lowest eigenspace of $H_0$. Indeed, 
\begin{equation}
\label{E:zero}
H_0 |{\cal N}_j\rangle = E_0 |{\cal N}_j\rangle,\quad E_0=\frac{U}{2}  {\cal N}({\cal N}-1)
\end{equation}
As to the first-order perturbative correction to  $|\Phi_j\rangle$, it  is standardly obtained as $|\Phi_j\rangle_1= (H_0 -E_0)^{-1} V |{\cal N}_j\rangle$.
After some simple algebraic manipulations  we get
\begin{equation}
\label{E:firstB}
|\Phi_j\rangle_1= \tau \sqrt{\cal N} \left( |\eta_j^+\rangle+ |\eta_j^-\rangle \right)
\end{equation}
where 
\begin{equation}
\label{E:firstC}
|\eta_j^+\rangle = \frac{(a^\dag_j)^{{\cal N}-1}a^\dag_{j+1}}{\sqrt{({\cal N}-1)!}} |\Omega\rangle \quad|\eta_j^-\rangle = \frac{(a^\dag_j)^{{\cal N}-1}a^\dag_{j-1}}{\sqrt{{(\cal N}-1)!}} |\Omega\rangle. \nonumber
\end{equation}
Hence, to the first order in $\tau$,
\begin{equation}
\label{E:qls}
|\Phi_j\rangle\simeq |{\cal N}_j\rangle + \tau \sqrt{\cal N} \left( |\eta_j^+\rangle+ |\eta_j^-\rangle \right)
\end{equation}
On the other hand it is easy to show that, to the first perturbative order in $\tau$, the solution to the semiclassical equations (\ref{E:fpe}) localized at lattice site $j$ is given by
\begin{equation}
\label{E:pert_class}
\phi_m \simeq \sqrt{\cal N} \delta_{m,j} + \tau \sqrt{\cal N}
(\delta_{m,j+1}+\delta_{m,j-1}).
\end{equation}
Plugging this expression into Eq.~(\ref{E:twf}) gives
\begin{equation}
\label{E:cls}
|\Phi\rangle\simeq |{\cal N}_j\rangle + \tau \sqrt{\cal N} \left( |\eta_j^+\rangle+ |\eta_j^-\rangle \right),
\end{equation}
i.e. the same result as in Eq.~(\ref{E:qls}).
Hence, to the first perturbative order,  the quantum superposition of the $M$ lowest eigenstates of $H$ defined in Eq.~(\ref{E:coh_sup}) coincides with the state obtained plugging the suitable localized solution to the semiclassical equations (\ref{E:fpe}) into the coherent state (\ref{E:twf}). We emphasize that in the strong interaction limit the zero temperature properties of $H$ can be investigated based on the ground state of the corresponding semiclassical model, Eq.~(\ref{E:scBH}), which is evaluated at a considerably lower computational cost.


\begin{thebibliography}{10}
\bibitem{N:Kanamoto2}
More precise estimates can be obtained considering the analytical solution to
  the NLSE on a ring domain \cite{N:Kanamoto}. Note that in this continuous
  model an "infrared" threshold related to the finite size of the system is
  naturally present. Conversely, the "ultraviolet" crossover related to the
  discreteness of the lattice has to be estimated requiring that $\Delta
  \approx 1$.
\bibitem{N:phase2}
This is obtained setting $\gamma_k = \langle \psi_1^{(k)}|f \rangle/|\langle
  \psi_1^{(k)}|f \rangle|$ where $|f\rangle$ is a state with a non vanishing
  projection on all of the $|\psi_1^{(k)}\rangle$. A suitable choice is
  $|f\rangle=|{\cal N}_m\rangle$. Note that $|\tilde \psi_1^{(k)}\rangle =
  \gamma_k |\psi_1^{(k)}\rangle$ is perfectly equivalent to
  $|\psi_1^{(k)}\rangle$, as an eigenstate of $H$. However, the $|\tilde
  \psi_1^{(k)}\rangle$'s are in some sense phase coherent, in that their phases
  are all aligned to that of the reference state $|f\rangle$, i.e. $\langle
  f|\tilde \psi_1^{(k)}\rangle = |\langle f|\tilde \psi_1^{(k)}\rangle|$.
\bibitem{N:Jack}
In Ref.~\cite{A:Jack} all of the quantum observables are plotted vs. $U/T =
  [({\cal N}-1) \tau]^{-1}$. Here we prefer to adopt $\tau$, which makes easier
  to put into evidence the scaling properties of the plotted quantities. Also
  note that our definition of $C_1$ differs from that in Ref.~\cite{A:Jack} by
  an irrelevant yet graphically convenient normalization factor.
\end{thebibliography}

\end{document}